\documentclass[journal]{IEEEtran}
\usepackage {graphicx}
\usepackage{setspace}
\usepackage{amssymb}
\usepackage{amsfonts}
\usepackage{mathrsfs}
\usepackage{amssymb,amsmath}
\usepackage{epstopdf}
\usepackage{bm}
\usepackage{algorithm}
\usepackage{algorithmic}
\usepackage{amsmath,amssymb,epsfig,graphics,subfigure}
\usepackage{flushend}
\usepackage{theorem}
\usepackage{array,color}
\usepackage[compress]{cite}
\usepackage{stfloats}
\usepackage{graphicx}
\usepackage{subfigure}
\usepackage{flushend}
\usepackage{array}
\usepackage{cases}
\usepackage{multirow}
\usepackage{chngpage}
\usepackage{longtable}
\usepackage{lineno}
\usepackage{amsfonts}
\usepackage{mathrsfs}
\usepackage{amssymb,amsmath}
\usepackage{algorithm}
\usepackage{algorithmic}
\usepackage{amsmath,amssymb,epsfig,graphics,subfigure}
\usepackage{theorem}
\usepackage{array,color}
\usepackage{amssymb}
\newcommand{\tb}{\mathbf}

\theoremheaderfont{\normalfont\bfseries}
\begin{document}
\title{Integrated Sensing and Communication-assisted Orthogonal Time Frequency Space Transmission for Vehicular Networks}
\author{Weijie Yuan,~\IEEEmembership{Member,~IEEE}, Zhiqiang Wei,~\IEEEmembership{Member,~IEEE}, Shuangyang Li,~\IEEEmembership{Student Member,~IEEE}, \\Jinhong Yuan,~\IEEEmembership{Fellow,~IEEE,} and Derrick Wing Kwan Ng,~\IEEEmembership{Fellow,~IEEE}
\thanks{
The authors are with the School of Electrical Engineering and Telecommunications, University of New South Wales, NSW 2052, Australia (e-mail: weijie.yuan@unsw.edu.au, zhiqiang.wei@unsw.edu.au, shuangyang.li@unsw.edu.au, j.yuan@unsw.edu.au, w.k.ng@unsw.edu.au).}
}
\maketitle
\begin{abstract}
Orthogonal time frequency space (OTFS) modulation is a promising candidate for supporting reliable information transmission in high-mobility vehicular networks. In this paper, we consider the employment of the integrated (radar) sensing and communication (ISAC) technique for assisting OTFS transmission in both uplink and downlink vehicular communication systems. Benefiting from the OTFS-ISAC signals, the roadside unit (RSU) is capable of simultaneously transmitting downlink information to the vehicles and estimating the sensing parameters of vehicles, e.g., locations and speeds, based on the reflected echoes. Then, relying on the estimated kinematic parameters of vehicles, the RSU can construct the topology of the vehicular network that enables the prediction of the vehicle states in the following time instant. Consequently, the RSU can effectively formulate the transmit downlink beamformers according to the predicted parameters to counteract the channel adversity such that the vehicles can directly detect the information without the need of performing channel estimation. As for the uplink transmission, the RSU can infer the delays and Dopplers associated with different channel paths based on the aforementioned dynamic topology of the vehicular network. Thus, inserting guard space as in conventional methods are not needed for uplink channel estimation which removes the required training overhead. Finally, an efficient uplink detector is proposed by taking into account the channel estimation uncertainty. Through numerical simulations, we demonstrate the benefits of the proposed ISAC-assisted OTFS transmission scheme.
\end{abstract}


\section{Introduction}
The automotive industry has developed significantly in the past decades to pave the way for the development of digital cities. With the new era of the fifth-generation (5G) mobile networks and beyond, it is expected that the vehicular communication will play an increasingly important role in people's daily lives, as it can support various promising applications, e.g., autonomous driving, traffic management, and on-the-go Internet services in intelligent transportation systems (ITS)\cite{siegel2017survey,wong2017key,9076668}. In addition to the requirement of efficient communication, highly accurate sensing is also required in vehicular networks, which is usually associated with radar systems. In particular, sensing the states, e.g., locations and speeds of vehicles as well as other objects is of great importance to realize collision-prevention and practical real-time road safety-related applications \cite{hult2016coordination,wang2017overview,lu2014connected}.

In the past, researches on radar sensing and communication were on two parallel streams since the frequency bands allocated for sensing and communication are usually separated for simplicity. Therefore, both systems rarely interfere with each other even they share similar underline signal processing functions and/or radio frequency (RF) units. Nevertheless, this conventional separated approach always underutilizes the system resources and is not sustainable for the development of long-term vehicular networks. In fact, both communication and radar communities seek more frequency spectrums to enable gigabit-level communication rate and to realize the massfication of automotive radars, respectively. 
Therefore, in vehicular networks where communication and sensing functionalities are highly demanded simultaneously, there is a trend to integrate both functionalities in a single system, a.k.a., integrated sensing and communication (ISAC), which allows the share of signal processing algorithms, hardware architectures, and frequency spectrum between communication and sensing systems \cite{zhang2020perceptive,rahman2020enabling}. By doing so, the cost in terms of hardware and spectrum resource can be reduced substantially while a higher system throughput can be achieved \cite{chiriyath2017radar,liu2020joint,gameiro2018research}.

Existing contributions on the topic of ISAC can be categorized into resource-sharing and common-wave approaches. The former splits the transmission resources into two orthogonal parts for supporting sensing and communication, respectively \cite{gonzalez2016radar}. In contrast, \cite{khawar2015coexistence,liu2018toward} proposed to use the same signal waveform to facilitate ISAC. By jointly designing the waveform for both functionalities, significant gains in terms of communication and sensing performance can be achieved, compared with the conventional resource-sharing approach. Also, motivated by the recent development of multi-carrier waveform-based radar sensing systems \cite{gogineni2013multi}, the orthogonal frequency division multiplexing (OFDM) technique \cite{shi2017power} has attracted numerous attentions for realizing ISAC. In particular, OFDM offers various advantages for communication purposes, such as resilience to multi-path fading, simple time and frame synchronization, and low-complexity data detection. On the other hand, for radar sensing, OFDM-based signals can also provide a satisfactory performance for target detection \cite{sen2010adaptive}. Various examples for OFDM-ISAC can be found widely in the literature \cite{sturm2011waveform,chiriyath2017radar}, where OFDM waveforms are adopted for both bi-static communication and mono-static radar \cite{chiriyath2017radar}. Although OFDM-ISAC systems has shown their advantages in some scenarios, the common problems of high peak-to-average-power ratio (PAPR) and vulnerability to time-varying channels still remain. In fact, it is worth noting that vehicular communication environments are inherently high-mobility e.g., $120$-$200$ km/h. In such scenarios, the orthogonality between the subcarriers in OFDM modulation no longer holds, which results in a dramatical performance degradation. This emerging need has recently motivated research interests in seeking for a new modulation scheme that is resilient to high-mobility channels.

Orthogonal time frequency space (OTFS), invented by Hadani \emph{et. al} \cite{7925924}, which modulates data symbols in the delay-Doppler (DD) domain rather than the conventional time-frequency (TF) domain, has shown its great potentials for providing high resilience against time-varying channels \cite{weiotfs,9321356}. In particular, it has been shown in the literature \cite{raviteja2018interference} that OTFS system significantly outperforms OFDM in terms of error performance in time-varying channels. By leveraging this two dimensional (2D) representation, each data symbol is spread over the entire TF domain, providing the chance to achieve the full TF diversity. Moreover, the set of DD domain basis functions is properly designed to combat the time and frequency selective fadings \cite{liotfs}. Also, OTFS can have a lower PAPR compared to that of OFDM, which allows the applications of highly efficient nonlinear power amplifiers. Overall, OTFS enjoys various advantages in high-mobility scenarios, which make it a promising candidate for vehicular communications. On top of that, OTFS modulation allows the direct interaction between the transmitted signals and the DD domain channel characteristics, e.g., delay and Doppler shifts, which aligns perfectly with the purpose of sensing. To this end, the pioneering work on OTFS-ISAC systems was reported in \cite{9109735}, which considered a single-antenna setting. {To realize the dual goal of sensing and communication, a maximum likelihood (ML) estimator was designed for estimating the sensing parameters from the echoes at the RSU, while message passing algorithm (MPA)-based receivers were adopted for data detection at the vehicles. 
On the other hand, OTFS-based radar sensing system with multiple-input multiple-output (MIMO) was developed in \cite{mariotfs}, where a hybrid digital-analog beamforming design was devised for parameter estimation and tracking. Both works have demonstrated the effectiveness of OTFS signaling for communication and sensing.
Nevertheless, the sensing parameters were not fully exploited for supporting the communication functionality. }

To serve multiple vehicles in the network, the RSU is generally equipped with  a massive antenna array to formulate multiple spatial beams \cite{chen2017vehicle}. However, a narrow beamwidth and high-mobility environment impose challenges on achieving an accurate beam alignment for communication. To reliably support the downlink communication, classic pilot-based beam tracking methods \cite{shaham2019fast, zhang2019codebook} adopted a few pilots to obtain the estimates of the angles relative to the RSU as well as the corresponding channel path loss factors and phase shifts. This information is then fed back to the RSU for formulating the transmit beamformers. Yet, this feedback-based protocol may result in exceedingly large amount of signaling  overhead and long latency. In contrast, benefiting from ISAC signals, the RSU can extract the angular parameters from the reflected echoes \cite{9171304,9246715}, which can be exploited by the RSU to predict the angles in the following time instant. Hence, the RSU can formulate its transmit beamformers to establish the communication links before transmitting the ISAC signals. 
As for the uplink communication, the uplink channel consists of a number of propagation paths when the vehicles are equipped with single antennas. To accurately estimate the multi-path channel, the conventional OTFS channel estimation approach \cite{8671740} adopted a single-pilot and inserted some guard space to avoid pilot contamination. The utilization of guard space incurs an overhead, which could be significant especially for vehicular communications requiring a short frame duration. On the other hand, after obtaining the channel estimates, an efficient detection algorithm is demanded for extracting the uplink information. Various detectors have been proposed in the literature, including the message passing algorithm \cite{li2020hybrid}, the variational Bayes \cite{9082873}, and the approximate message passing \cite{yuan2020iterative}, which allow low-complexity implementations at the cost of moderate bit-error-rate (BER) performance degradation. However, all aforementioned detectors rely on the availability of perfect channel state information (CSI), which is overly optimistic. More importantly, applying their results may suffer from performance loss in practice.

In this paper, we aim for designing a novel ISAC-assisted OTFS transmission scheme in vehicle-to-infrastructure (V2I) scenarios, where both uplink and downlink transmissions are considered. Exploiting the OTFS-ISAC signals, the RSU can obtain the estimates of delays, Dopplers, and angles associated with vehicles in the communication range, which can be adopted for constructing the dynamic topology of the vehicular network. {Motivated by the slow time-varying property of the DD domain channel, the RSU can predict the channel parameters and perform pre-equalization to combat channel dynamics for the downlink transmission, which is sufficient for achieving a good error performance at the vehicles. }
As for the uplink transmission, we can predict the delays and Dopplers associated with different paths at the RSU in a similar manner. Given the predicted interference pattern, we design a new symbol placement scheme to facilitate the channel estimation. In particular, the pilot and data symbols are superimposed while the guard space can be absent. Accordingly, the entire uplink OTFS frame can be used for carrying information, which significantly reduces the required training overhead. To improve the uplink performance, we further develop a factor graph \cite{kschischang2001factor} and sum-product algorithm (SPA)-based detector, where the message derivation takes into account the channel estimation uncertainty. Our simulation results show that the proposed algorithm can accurately track the angular parameters for transmit beamforming design and can reliably support the downlink communications. Moreover, for the uplink transmission, the detection performance of the proposed algorithm can approach that of the ideal case with perfect CSI. {Our contributions are summarized as follows:}
\begin{itemize}
  \item We develop an ISAC-assisted OTFS transmission scheme for vehicular networks. The OTFS signals are used for both sensing the vehicle states and downlink communication, which helps to reduce the hardware cost as well as the spectral resources.
  \item We propose a predicted transmit beamforming design for the downlink transmission based on the estimated kinematic parameters of vehicles from the reflected echoes. The effect of the downlink channel is compensated at the RSU which allows the use of direct single-tap ML detectors at the vehicles without the need of channel estimation.
  \item We propose to predict the delay and Doppler shifts associated with the paths by exploiting the slow time-varying property of the DD domain channel. Based on the prediction of the channel, we design a new symbol placement scheme that does not require the guard space as needed in conventional schemes, which significantly reduces the training overhead. Furthermore, we propose a factor graph and the SPA-based detector while taking the channel estimation uncertainty into account.
\end{itemize}

\emph{Notations:}
Unless otherwise specified, we use a boldface capital letter, a boldface lowercase letter, and a calligraphy letter to denote a vector, a matrix, and a set, respectively; the superscripts $\textrm{T}$, $*$, {and} $\textrm{H}$ denote the transpose, conjugate, and {the} Hermitian operations{, respectively};
%
$\delta(\cdot)$ denotes the Dirac delta function; $|\cdot|$ denotes the modulus of a complex number or the cardinality of a set; $\|\cdot\|$ denotes the $\ell^2$ norm; $[\cdot]_N$ denotes the modulo operation with divisor $N$; $\mathcal{X}\sim p$ denotes all variables in set $\mathcal{X}$ except the $p$-th entry; $\mathbb{C}$ denotes a complex space; $\hat{x}$ denotes the estimation of variable $x$ based on collected measurements and $\bar{x}$ denotes the prediction of variable $x$ based on the state evolution; the plain font subscripts $*^{\rm D}$ and $*^{\rm U}$  refers to the downlink and uplink information, respectively; $\propto$ indicates that the left side is proportional to the right side; $\arcsin$ denotes the inverse of the sine function; $\delta(\cdot)$ denotes the Dirac delta function; $\mathbb{E}$ denotes the expectation operator; $\mathbb{V}$ denotes the operator to determine the variance. As there are different channels and transmitted/received signals, for clarity, we further summarize the notations of some commonly used variables in Table \ref{table11}.

\begin{table}[]
\centering
{
\caption{List of Variable Notations}
\label{table11}
\begin{tabular}{|l||l|}
\hline
$s_i(t)$& Downlink ISAC signal to vehicle $i$\\
$\tb{H}(t,\tau)$ & Radar sensing channel\\
$\tb{h}_i(t)$ & Downlink communication channel to vehicle $i$\\
$\tb{h}^{\rm U}_i(t,\tau)$ & Uplink communication channel for vehicle $i$\\
$\tb{r}$ & Received radar echoes\\
$y_i$ & Downlink received signal at vehicle $i$\\
$y_i^{\rm U}$ &Uplink received signal of vehicle $i$ at the RSU\\
$\tb{z}$ & Sensing noise term \\
$w$ & Communication additive noise\\
\hline
\end{tabular}
}
\end{table}

\section{System Model}
\begin{figure}[!h]
\centering
\includegraphics[width=.5\textwidth]{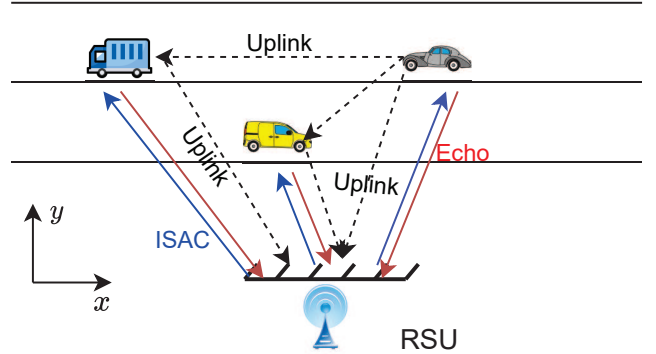}
\caption{{The considered vehicular network. The blue solid lines denote the downlink ISAC signals. The red solid lines denote the echos reflected by vehicles. The black dashed lines denote the uplink communication signals.}}%
\label{network}%
\end{figure}

As shown in Fig. \ref{network}, let us consider a network with $P$ vehicles supported by a roadside unit (RSU). The RSU is equipped with a transmit uniform linear array (ULA) of $N_t$ antennas and a separate receive ULA of $N_r$ antennas. Under the assumption of sufficient isolation between the transmit and receive arrays, the radar echoes would not interfere with the downlink transmissions \cite{palacios2019hybrid}. The ULAs of the RSU are parallel with the road, hence the angle-of-arrival (AoA) is identical to the angle-of-departure (AoD). For the vehicles, we consider a point target model and assume that they are equipped with a single-antenna for receiving downlink information from the RSU and transmit their uplink information to the RSU. Without loss of generality, we set the road as the $x$-axis, the vehicles only moves along the positive or negative directions of the $x$-axis.

\subsection{OTFS Modulation}
In this paper, we consider OTFS modulation for both the uplink and downlink transmissions. As typically discussed in the OTFS literature, the sequence of information bits sent to vehicle $i$ is mapped to $N\times M$ data symbols, which are placed in the 2D delay-Doppler (DD) domain, denoted by $x_i\left[k,l\right]$ where $0\leq l\leq M-1$ and $0\leq k\leq N-1$ refers to the delay and Doppler indices, respectively. Integers $N$ and $M$ indicates the number of time slots and the number of subcarriers for the OTFS frame, respectively. The DD domain symbols are then spread over the time-frequency (TF) domain using the inverse symplectic finite Fourier transform (ISFFT), given by \cite{7925924}
\begin{align}\label{ddtotf}
X_i\left[n,m\right] = \frac{1}{\sqrt{MN}}\sum_{k=0}^{N-1}\sum_{l=0}^{M-1} x_i\left[k,l\right]e^{j2\pi \left(\frac{nk}{N}-\frac{ml}{M}\right)},
\end{align}
where $n$ and $m$ denotes the time and frequency indices, respectively. Then, the RSU generates the transmitted ISAC signal in the time domain as \cite{raviteja2018interference}
\begin{align}\label{tftot}
s_i(t)=\sum_{m=0}^{M-1}\sum_{n=0}^{N-1}X_i\left[n,m\right]e^{j2\pi m \Delta_f (t-nT)} g_{\rm tx}(t-nT),
\end{align}
where $g_{\rm tx}(t)$ is the transmit filter, $T$ and $\Delta f$ denote the symbol duration and frequency spacing, which satisfy $T\Delta f = 1$ to maintain the orthogonality.

\subsection{Radar Signal Model}
To support all $P$ vehicles, the RSU first prepares a multi-beam ISAC signal $\tb{s}(t)$ with $P$ dimensions\cite{zhang2018multibeam}, i.e.,
\begin{align}
	\tb{s}(t) = \left[s_1(t),...,s_{P}(t)\right]^{\rm T},
\end{align}
where the $i$-th signal $s_i(t)$ carries the information to vehicle $i$.

Next, the signal $\tb{s}(t)$ is transmitted over all $N_t$ antennas using a beamforming matrix $\tb{F}$, formulating as
\begin{align}
\tilde{\tb{s}}(t) = \tb{F} \tb{s}(t),
\end{align}
where $\tb{F}\in\mathbb{C}^{N_t\times P}$ is used for steering the beams to the intended directions, whose $i$-th column is given by
\begin{align}\label{bf_vector}
\tb{f}_{i} = \sqrt{\frac{p_i}{N_t}}\tb{a}(\theta_i),
\end{align}
which is used to steer the transmitted signal $s_i(t)$ towards the intended direction $\theta_i$. In \eqref{bf_vector}, $p_i$ denotes the allocated transmit power to vehicle $i$ and $\tb{a}(\theta_i)$ denotes the steering vector $\tb{a}(\theta_i) = \left[a_1(\theta_i),...,a_{N_t}(\theta_i)\right]^{\rm T}$ with
\begin{align}
	a_n(\theta_i) = e^{j (n-1)\pi \sin\theta_i}.
\end{align}
The intended direction $\theta_i$ is unknown and thus is usually set as the predicted angle of the $i$-th vehicle relative to the RSU, denoted by $\bar{\theta}_i$, to achieve the desired beamforming gain.

The transmitted ISAC signal is then reflected by $P$ moving vehicles and the echo is received at the radar detector mounted at the RSU. From the view of the RSU, the radar sensing channel is both time and frequency selective, which is given by
\begin{align}
\tb{H}(t,\tau)  = \sum_{i=1}^P \beta_i \tb{b}(\theta_i) \tb{a}^{\rm H}(\theta_i) \delta(\tau-\gamma_i) e^{j2\pi \omega_i t},
\end{align}
where 
$\tb{b}(\theta_i)= \left[b_1(\theta_i),...,b_{N_r}(\theta_i)\right]^{\rm T}$ is the receive steering vector, satisfying $b_n(\theta_i) = a_n(\theta_i)$, and $\theta_i$ is the angle of the $i$-th vehicle relative to the RSU. The terms $\beta_i$, $\gamma_i$, and $\omega_i$ denote the reflection coefficient, the round-trip delay, and the round-trip Doppler spread corresponding to the $i$-th vehicle, respectively. Finally, the $P$-dimensional received echo can be written by
\begin{align}
\tb{r}(t) = \sum_{i=1}^P \beta_i \tb{b}(\theta_i) \tb{a}^{\rm H}(\theta_i) \tilde{\tb{s}}(t-\gamma_i)e^{j2\pi \omega_i t} + \tb{z}(t),
\end{align}	
where $\tb{z}(t)\in\mathbb{C}^{N_r\times 1}$ is the additive white Gaussian noise process.

\subsection{Radar Measurement Model}
Given the massive MIMO scenario and the favourable channel condition, the receive steering vectors $\tb{b}(\theta_i)$ corresponding to different vehicles are asymptotically orthogonal \cite{ngo2015massive}, following
\begin{align}
	\tb{b}^{\rm H}(\theta_i) \tb{b}(\theta_{i'})\to 0,~\forall~i\neq i',~N_r\to \infty.
\end{align}
Therefore, the RSU can distinguish the echoes reflected by different vehicles, i.e., $\tb{r}(t)=[r_1(t),...,r_P(t)]^{\rm T}$. After acquiring the echoes of all vehicles, the RSU can perform radar's matched filtering by using a bank of transmitted signals $s_i(t),~1\leq n\leq P$ with delay $\gamma$ and Doppler $\omega$, formulating as
\begin{align}
&\tilde{\tb{r}}_i(\gamma,\omega) = \int_0^{\Delta T} {r}_i(t) s^{*}_i(t-\gamma)e^{-j\omega t}\textrm{d} t\\
& =  \int_0^{\Delta T} \beta_i \tb{b}(\theta_i) \tb{a}^{\rm H}(\theta_i)  \tilde{\tb{s}}(t-\gamma_i)s^{*}_i(t-\gamma)e^{-j(\omega-\omega_i) t}\textrm{d} t+ {\tb{z}}_i,\nonumber
\end{align}
where $\Delta T = N\cdot T$ denotes the OTFS frame duration and $\tb{z}_i = \int_0^{\Delta T}\tb{z}(t)s^{*}_i(t-\gamma)e^{-j\omega t}\textrm{d} t$ is the filtered noise vector, whose elements are assumed independent and identical Gaussian distributed with zero mean and variance $\sigma^2$. A peak occurs at the matched filter output when $\gamma$ and $\omega$ are perfectly tuned to the corresponding delay and Doppler shifts of the $i$-th vehicle \cite{friedlander2012transmit}, respectively. Therefore, after radar's matched filtering, the estimates of the signaling delays and Doppler shifts corresponding to the vehicles can be obtained, denoted by $\hat{\gamma}_i$ and $\hat{\omega}_i$, $\forall~i$. Having obtained the delay and Doppler estimates, the received signal can be written as
\begin{align}\label{observation_r}
\tb{r}_i = {G_m} \beta_i\tb{b}(\theta_i) \tb{a}^{\rm H}(\theta_i)  \tb{f}_i+\tb{z}_i,
\end{align}
where $G_m$ is the signal-to-noise ratio (SNR) gain obtained by radar's matched filtering, which is in general identical to the energy of the signal $s_i(t)$.

 \subsection{Communication Model}
Then, we elaborate the downlink communication channel model. Providing the asymptotical orthogonality of the steering vectors to different directions\cite{ngo2015massive}, the downlink channel can be modeled by a LoS-dominated one, i.e.,
 \begin{align}
 	\tb{h}^{\rm D}_i(t) = \sqrt{\frac{c}{4\pi f_c d^2_i}} \tb{a}^{\rm H}(\theta_i)  e^{j2\pi \nu_{i}t},
 \end{align}
 where $c$ is the speed of light, $f_c$ is carrier frequency, $\nu_i$ is the Doppler shift due to the movement of the $i$-th vehicle, and $d_i$ is distance between the $i$-th vehicle and the RSU\footnote{It should be noted that $\gamma_i$ and $\omega_i$ are the delay and Doppler shifts associated with the radar echoes while $\tau_i$ and $\nu_i$ are associated with the downlink communication signals. As $\omega_i$ is the round-trip Doppler shift, it is in general $2\nu_i$.}.
Consequently, the received signal at the $i$-th vehicle is given by
 \begin{align}
{y}_i(t) = 	h_{i} e^{j2\pi \nu_{i}t} \tb{a}^{\rm H}(\theta_i)  \tb{f}_i s_i(t) + w(t),
 \end{align}
where $w(t)$ denotes the additional Gaussian noise with power spectral density (PSD) $N_0$ and we use the shorthand notation $h_i = \sqrt{\frac{c}{4\pi f_c d_i^2}}$.
Next, the multi-carrier demodulation and receive filtering $g_{\rm rx}(t)$ are performed to obtain the TF domain samples, i.e.,
\begin{align}\label{TF}
Y_i\left[n,m\right] = \int y_i(t) g_{\rm rx}^{*}(t-n\Delta T)e^{-j 2 \pi m \Delta f(t-nT)} \textrm{d} t.
\end{align}
Finally, the TF domain samples are transformed to the DD domain via symplectic finite Fourier transform (SFFT):
\begin{equation}\label{DD_in_out}
y_i\left[ {k,l} \right] = \frac{1}{{\sqrt {NM} }}\sum\limits_{n = 0}^{N - 1} {\sum\limits_{m = 0}^{M - 1} {Y_i\left[ {n,m} \right]{e^{-j2\pi \left( {\frac{{kn}}{N} - \frac{{lm}}{M}} \right)}}} }.
\end{equation}
For simplicity, we consider a sufficient Doppler resolution such that $\nu_i = \frac{k_i}{N T}$, where $k_i$ is an integer \cite{raviteja2018interference}. Invoking ideal transmit and receive filterings and after some manipulations, we obtain the DD domain input-output relationship for the downlink transmission as
\begin{align}
	y_i\left[ {k,l} \right]  = h_i \tb{a}^{\rm H}(\theta_i) \tb{f}_i x_i\left[ {(k-k_i)_N,(l-l_i)_M} \right] + w\left[ {k,l} \right],
\end{align}
where $l_i=0$, $w\left[ {k,l} \right]$ is the additive noise sample, having the PSD of $N_0$.

Next, we consider the uplink vehicle-to-RSU transmission model. Considering that the vehicle is equipped with a single-antenna, the transmitted signal will be scattered by other vehicles in the network. Therefore, the multi-path channel model be written as\footnote{For ease of exposition, here we follow the general assumption \cite{mariotfs} that the number of scatterings in the environment is identical to the number of vehicles.}
 \begin{align}
 	\tb{h}^{\rm U}_i(t,\tau) = \sum_{p=1}^P h_{i,p}   \tb{b}(\theta_{i,p}) \delta(\tau-\tau_{i,p}) e^{j2\pi \nu_{i,p} t},
 \end{align}
where $h_{i,p}$, $\tau_{i,p}$, $\nu_{i,p}$, and $\theta_{i,p}$ denote the channel gain, the delay, the Doppler, and the angle relative to the RSU for the $p$-th path of vehicle $i$, respectively. Specifically, the first path $p=1$ corresponds to the direct path from the $i$-th vehicle to the RSU, having a delay of $\tau_{i,1} = 0$. By defining the uplink DD domain transmitted symbols by $\tb{x}_i^{\rm U} = \{x^{\rm U}_i [k,l]\},~0\leq k \leq N-1, ~0\leq l \leq M-1$, following the OTFS modulation process in Sec. II-B, the uplink transmitted signal is denoted by $s^{\rm U}_i(t)$. To receive the signal sent by the $i$-th vehicle, the RSU adopts a bank of receive beamfomers $\tb{g}_p\in\mathbb{C}^{N_r\times 1}$ and the received signal can be written as
\begin{align}
	y^{\rm U}_i(t)  =  \sum_{p=1}^{P} h_{i,p} \tb{g}^{\rm H}_p \tb{b}(\theta_{i,p})s_i(t-\tau_{i,p}) e^{j2\pi \nu_{i,p} t} + \tilde{w}(t).
\end{align}

Similar to the OTFS demodulation for the downlink transmission, after multi-carrier demodulation, receive filtering, and SFFT, we arrive at the input-output relationship for the DD domain received signal, formulating as
\begin{align}\label{io_dd_up}
y_i^{\rm U}\left[k,l\right] =  \sqrt{N_r} \sum_{p=1}^{P}& h_{i,p} \tb{g}^{\rm H}_p \tb{b}(\theta_{i,p}) e^{-j2\pi \nu_{i,p}\tau_{i,p}} \\\cdot&x_i^{\rm U}\left[(k-k_{i,p})_N,(k-l_{i,p})_M\right] + \tilde{w}\left[k,l\right],\nonumber
\end{align}
where $k_{i,p} = \frac{\nu_{i,p}}{N T}$ and $l_{i,p} = \frac{\tau_{i,p}}{M \Delta f}$, and $\tilde{w}\left[k,l\right]$ denote the Doppler index, the delay index, and the DD domain noise sample, respectively. Equation \eqref{io_dd_up} can also be expressed as a 2D convolution of the transmitted symbols and the DD domain effective channel, i.e.,
\begin{align}\label{io_dd}
y_i^{\rm U}\left[ {k,l} \right]=&\sum\limits_{k' = 0}^{N - 1} {\sum\limits_{l' = 0}^{M - 1} {x_i^{\rm U}\left[ {k',l'} \right]} } {h_i^{\rm U}}\left[ {\left(k - k'\right)_N,\left(l - l'\right)_M} \right]\nonumber\\&+\tilde{w}\left[k,l\right],
\end{align}
where the effective channel ${h_i^{\rm U}}\left[k,l\right] $ is given by
\begin{align}
	{h_i^{\rm U}}\left[k,l\right] = \sum_{p=1}^{P} h_{i,p} \phi\left(k-k_{i,p},l-l_{i,p}\right)e^{-j2\pi {\tau_{i,p}}\nu_{i,p}},
\end{align}
with the DD domain filter expressed as
\begin{align}
&\phi\left(k-k_{i,p},l-l_{i,p}\right) \nonumber\\&= \frac{1}{MN}\sum_{n=0}^{N-1}\sum_{m=0}^{M-1}e^{-j2\pi n \frac{k-k_{i,p}}{N}}e^{-j2\pi m \frac{l-l_{i,p}}{M}}.
\end{align}

 \section{ISAC-assisted OTFS Communications}
In this section, we discuss the proposed ISAC-assisted OTFS transmission scheme. We will first summarize the proposed framework. Then, we will study the sensing parameter estimation and discuss the prediction of the communication channel parameters. Finally, receiver design for uplink and downlink transmissions will be investigated.

\subsection{General Framework for ISAC-assisted OTFS Communications}
 The proposed ISAC-assisted OTFS transmission scheme has the following steps to fulfill both communication and sensing functionalities.

 \textbf{1) State estimation:} At time instant $\eta$, the RSU sends the OTFS-ISAC signals to all vehicles. The reflected echoes are received at the RSU, which is used for estimating the motion parameters, i.e., delays, Dopplers, and angles of vehicles at instant $\eta$.

 \textbf{2) Dynamic topology construction and prediction:} The delays, Dopplers, and angles of vehicles can be adopted for inferring the locations and speeds of the vehicles and time $\eta$, which helps the RSU to construct the dynamic topology of the vehicular network. Then the RSU can predict the speeds and locations of the vehicles in the following time instant $\eta+1$.

 \textbf{3) Assistance to downlink communication:} Based on the predicted speeds and locations of the vehicles at time instant $\eta+1$, the RSU can also predict the angles and channel impairments associated with vehicles at time $\eta+1$. Then, the RSU can formulate its transmit beamformers and combat the channel impairments based on the predicted parameters before transmitting the OTFS-ISAC signals at time $\eta+1$.

 \textbf{4) Assistance to uplink communication:} Based on the predicted locations and speeds from Step 2, the RSU also obtains the relative distances and speeds between vehicles, which are converted to the delays and Dopplers associated with the uplink DD domain channel at time instant $\eta+1$. As the interference pattern of the multi-path channel is known to the RSU, new symbol placement scheme with much lower training overhead is proposed, which will be shown in Sec. III-D.

 \subsection{Sensing Parameter Estimation and Prediction}
 Based on the received echoes \eqref{observation_r}, the RSU is capable of inferring the angular parameters and the reflected coefficients. For the reflected coefficient $\beta_i$, it is calculated as
  \begin{align}
 \beta_{i} = \frac{\xi}{2 d_i},	
 \end{align}
 where $\xi$ represents the radar cross section (RCS) \cite{skolnik2001radar}. Considering that the delay of echo ${\gamma}_i$ can be expressed as the round-trip range distance dividing by the signal propagation speed, i.e., ${\gamma}_i =\frac{2 d_i}{c}$, and using the estimated delay $\hat{\gamma}_i$ obtained at the RSU, the estimate of $\beta_i$ can be easily obtained as $\hat{\beta}_i = \frac{\xi}{c\hat{\gamma}_i}$. Substituting $\hat{\beta}_i$ into \eqref{observation_r} and after straightforward manipulations, we obtain
 \begin{align}
{\tb{r}}_{i} = &\hat{\beta_i} G_m \sqrt{p_{i}} \left[
\begin{array}{c}
1 \\
e^{j\pi\sin\theta_{i}}\\
...\\
e^{j\pi(N_r-1)\sin\theta_{i}}
\end{array}
\right]\sum\limits_{i=1}^{N_t}e^{j\pi(i-1)(\sin\bar{\theta}_{i}-\sin\theta_{i})}\nonumber\\&+{\tb{z}}_{i}.
 \end{align}

 From the perspective of the optimal ML estimation, we aim at maximizing the likelihood function $p(\tb{r}_i|\theta_i)$. Given the number of the transmit antennas $N_t$, we can represent the possible values of $\theta_i$ by a discrete set of $N_t$ angles, denoted by $\bm{\Theta}=\{\frac{\pi}{N_t},\frac{2\pi}{N_t},...,{\pi}\}$. Therefore, the maximization of the likelihood function is given by
 \begin{align}\label{ML_radar}
 \hat{\theta}_i = \arg\max_{\theta_i\in\bm{\Theta}} 	p(\tb{r}_i|\theta_i).
 \end{align}
Observed from \eqref{ML_radar}, for the angle relative to each vehicle, an exhausted search in the set $\bm{\Theta}$ is required to find the optimal estimate. To reduce the estimation complexity, we could only choose the values in the set $\bm{\Theta}$ that are close to the predicted angle $\bar{\theta}_i$ to find the estimated angle $\hat{\theta}_i$ that maximizes the likelihood function $p(\tb{r}_i|\theta_i)$.

Having obtained the delay, the Doppler, and the angle estimates of all vehicles relative to the RSU, the RSU is capable of determining the estimated vehicle locations as well as their speeds in the network. Let us denote the Cartesian coordinates of the RSU and the $i$-th vehicle by $\tb{q}_{\rm RSU} = [q_{x,{\rm RSU}},q_{y,{\rm RSU}}]^{\rm T}$ and $\tb{q}_i = [q_{x,i},q_{x,i}]^{\rm T}$, where the subscripts $x$ and $y$ denote the coordinates on $x$-axis and $y$-axis, respectively. Given the estimates of the delay $\hat{\gamma}_i$ and the angle $\hat{\theta}_i$, we can estimate the location of vehicle $i$ at the current instant
\begin{align}
	\hat{q}_{x,i} = q_{x,{\rm RSU}} + \frac{c\hat{\gamma}_i\sin\hat{\theta}_i}{2},\\
	\hat{q}_{x,i} = q_{y,{\rm RSU}} + \frac{c\hat{\gamma}_i\cos\hat{\theta}_i}{2}.
\end{align}
Furthermore, the speed for vehicle $i$ can be estimated by using its relationship with the Doppler estimate, i.e.,
\begin{align}
\hat{s}_i = \frac{c \hat{\omega}_i}{f_c \cos \hat{\theta}_i}.  	
\end{align}
The estimated speed $\hat{s}_i$ may be positive or negative, which indicates the moving direction of vehicle $i$.

\begin{figure}[!t]
\centering
\includegraphics[width=.5\textwidth]{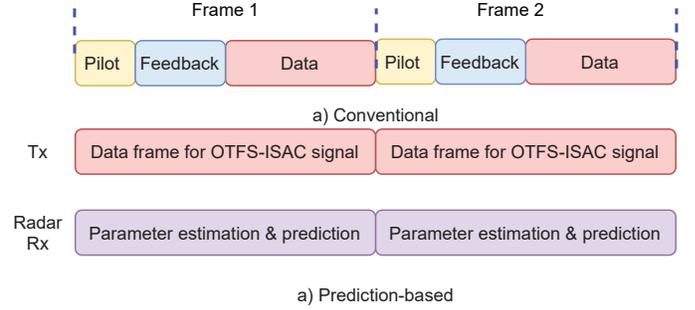}
\caption{{Comparison of the downlink communication protocols for the conventional and the prediction-based beam alignment schemes.}}%
\label{frame}%
\end{figure}

Based on the estimated vehicle locations and speeds, we can predict the motion parameters for vehicles in the following OTFS frame transmission. Specifically, for the OTFS frame duration of $\Delta T$ and the predicted location for the $i$-th vehicle in the following time instant can be expressed as
\begin{align}
	\bar{\tb{q}}_i = \hat{\tb{q}}_i + \Delta T \cdot \hat{s}_i\cdot \left[1,0\right]^{\rm T}.
\end{align}
Moreover, we are able to predict angle $\theta_i$ of the $i$-th vehicle relative to the RSU, given by
\begin{align}\label{predicted_angle}
\bar{\theta}_i^{\rm new} =  \arcsin \frac{\bar{q}_{x,i}-q_{x,{\rm RSU}}}{\|\bar{\tb{q}}_i -\tb{q}_{\rm RSU}\|}.
\end{align}
In this paper, we assume that the speeds of vehicles do not change in the relatively short time duration, i.e., $\bar{s}_i = \hat{s}_i$, as commonly adopted in the literature \cite{9171304}. Based on the predicted angle $\bar{\theta}_i^{\rm new}$, the RSU is able to formulate a transmit beamformer for the next time instant. In Fig. \ref{frame}, we depict the downlink communication protocols for the conventional beam alignment scheme and the prediction-based beam alignment schemes. It can be seen that for the conventional beam alignment scheme, the RSU has to transmit dedicated pilots to all vehicles to obtain the related angular parameters. Then the vehicles feed back the angle estimates to the RSU for formulating the transmit beamforming, followed by the data transmission. In contrast, for the proposed scheme, the angles are estimated based on the echoes of the ISAC signals, followed by a prediction procedure. Consequently, the RSU can steer the transmit array before the data transmission, which means the whole frame can be used for carrying data information. Apparently, the signaling latency and overhead can be reduced. Moreover, for the proposed OTFS-ISAC scheme, the RSU can exploit the whole frame as the pilots for estimating the angular parameters, which is expected to achieve a better estimation performance compared to the conventional scheme relying on only a few pilots.

\subsection{Downlink Communication}
As discussed in Sec II. D, the downlink communication channel is LoS dominated. In conventional downlink communication relying on OTFS modulation, we usually insert a pilot and reserve some guard space in the OTFS data frame for estimating the channel \cite{8671740}. However, with the sensing capability of the RSU, we can compensate the channel effect at the RSU side for downlink transmission. Thus, the vehicles can bypass the need of channel estimation for data detection.

 Given the prediction of the motion parameters of vehicles, the RSU could simply obtain the predicted channel gain and the Doppler corresponding to the $i$-th vehicle, which are given by
\begin{align}
\bar{h}_i &= \frac{c}{4\pi f_c \|\bar{\tb{q}}_i -\tb{q}_{\rm RSU}\|}~\textrm{and}\\
\bar{\nu}_i &= 	\frac{\bar{s}_i\cos\bar{\theta}_i f_c}{c},
\end{align}
respectively. Using $\bar{h}_i$ and $\bar{\nu}_i$, the RSU can pre-equalize the transmitted signal to compensate the channel attenuation at the transmitter side and the vehicles do not have to estimate the channel after acquiring the OTFS-ISAC signal sent by the RSU. In particular, after adopting pre-equalization, the received signal can be written as
	 \begin{align}
{y}_i(t) = 	\frac{h_{i}}{\bar{h}_i} \sqrt{p_i N_t} e^{j2\pi (\nu_{i}-\bar{\nu}_i)t} \tb{a}^{\rm H}(\theta_i)   \tb{a}(\bar{\theta}_i)  s_i(t) + w(t).
 \end{align}
Provided that the predicted channel gain and Doppler shift are sufficiently accurate, the DD domain received sample can be approximately written as
\begin{align}\label{pre_equal}
	y_i\left[ {k,l} \right]  \approx \sqrt{p N_t} \tb{a}^{\rm H}(\theta_i)\tb{a}(\bar{\theta}_i)x_i\left[ {k,l} \right] + w\left[ {k,l} \right].
\end{align}
A single-tap ML detector can be adopted for inferring the transmitted symbols, following
\begin{align}\label{ML_downlink}
x_i\left[ {k,l} \right]  = \arg\max_{x_i\in\mathcal{A}} \left|y_i\left[ {k,l} \right]  - \sqrt{p N_t} \tb{a}^{\rm H}(\theta_i)\tb{a}(\bar{\theta}_i)x_i\left[ {k,l} \right]\right|^2,
\end{align}
where $\mathcal{A}$ denotes the set of the constellation points set of the transmitted symbols. We can see that by exploiting the sensing parameters, the downlink communication channel parameters are predictable at the RSU side and all vehicles can directly detect the data symbols without channel estimation.

Moreover, the alignment of the beams can affect the communication performance. From \eqref{pre_equal}, the receive SNR at the $i$-th vehicle can be expressed as
\begin{align}
\textrm{SNR}_i&=	\frac{\left|\sqrt{p_i N_t} \tb{a}^{\rm H}(\theta_i)\tb{a}(\bar{\theta}_i)x_i\left[k,l\right]\right|^2}{N_0}\nonumber\\
&= p_i N_t E_s \frac{\left|\tb{a}^{\rm H}(\theta_i)\tb{a}(\bar{\theta}_i)\right|^2}{N_0},
\end{align}
where $E_s$ denotes the power of the data symbol. Obviously, when the predicted angle $\bar{\theta}_i$ matches the actual angle $\theta_i$, the highest antenna array gain as well as the maximum receive SNR are achieved.


\subsection{Uplink Communication}
Unlike the downlink communication scenario, the uplink transmission still involves multi-path propagations. Using the predicted locations $\bar{\tb{q}}_i$, the speeds $\bar{s}_i$, and the angles $\bar{\theta}_i$ for all vehicles, the RSU can predict the delay $\tau_{i,p}$ as well as the Doppler $\nu_{i,p}$ based on the geometric relationship of the vehicles.

For instance, the uplink signal sent from vehicle $i$ to the RSU is reflected by the $j$-th vehicle, which contributes to the $p$-th path of the multi-path channel. Consequently, we are capable of predicting the associated delay and Doppler of the $p$-th path, i.e.,
\begin{align}
\bar{\tau}_{i,p} &= \frac{\|\bar{\tb{q}}_i-\bar{\tb{q}}_j\|+\|\bar{\tb{q}}_j-{\tb{q}}_{\rm RSU}\|-\|\bar{\tb{q}}_i-{\tb{q}}_{\rm RSU}\|}{c},\\
\bar{\nu}_{i,p} & = \frac{\cos\bar{\theta}_jf_c}{c}\left(\bar{s}_i-\bar{s}_j\right),
\end{align}
respectively, where $\bar{\tb{q}}_j$, $\bar{s}_j$, and $\bar{\theta}_j$ denote the location, the speed, and the angle of vehicle $j$, respectively. For receive beamforming design, we can simply set $\tb{g}_p=\tb{b}(\bar{\theta}_{j})$ based on the prediction of the angle relative to the $j$-th vehicle. In particular, the Doppler associated with the direct path, i.e., $\bar{\nu}_{i,1}$ is the reverse of the downlink Doppler, given by $\bar{\nu}_{i,1}=-\bar{\nu}_i$. Given sufficient resolution for the delay and the Doppler, the interference pattern of the uplink channel is known to the RSU. Nevertheless, the channel gains of different paths are stochastic rather than deterministic. Even with the motion parameters of all vehicles, it is still unable to infer the channel gains of all paths as they are randomly distributed following the power delay profile. Therefore, channel estimation is a prerequisite for decoding the uplink information.

 \begin{figure}[!t]
\centering
\includegraphics[width=.48\textwidth]{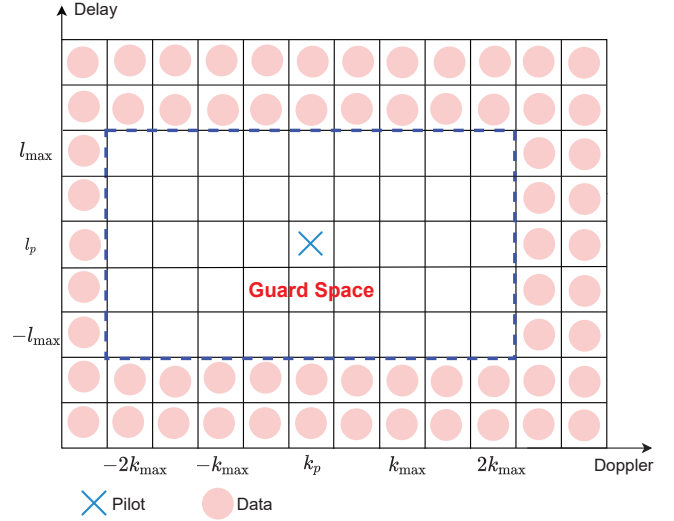}
\caption{{Symbol placement scheme in a single OTFS frame for the proposed channel estimation.}}%
\label{conventional}%
\end{figure}

\subsubsection{Uplink Channel Estimation}\label{subsubsec1}
By exploiting the 2D convolutional relationship of the transmitted symbols and the uplink DD domain channel, conventional channel estimation \cite{8671740} for OTFS modulation employs only one pilot at grid $[k_p,l_p]$ and insert a guard space for avoiding the interference between the pilot and data symbols, as illustrated in Fig. \ref{conventional}. The size of the guard space depends on the maximum delay and Doppler indices $l_{\rm max}$ and $k_{\rm max}$. After transmitting through the channel, the pilot will not spread out of the guard space. As such, one can easily estimate the channel by observing the received samples with indices $0\leq l\leq l_p+l_{\rm max}$ and $k_p-k_{\rm max}\leq l\leq k_p+k_{\rm max}$ and comparing them with a preset threshold\footnote{The threshold is set to identify whether the received samples are contributed by the pilot. A general value is $3\sqrt{N_0}$ for additive Gaussian white noise with power spectral density $N_0$ \cite{8671740}.}. Nevertheless, using guard space will inevitably deteriorate the communication efficiency, as the high-mobility vehicles require a larger size of guard space while the vehicular communication demands short duration for the OTFS frame.

 \begin{figure}[!t]
\centering
\includegraphics[width=.5\textwidth]{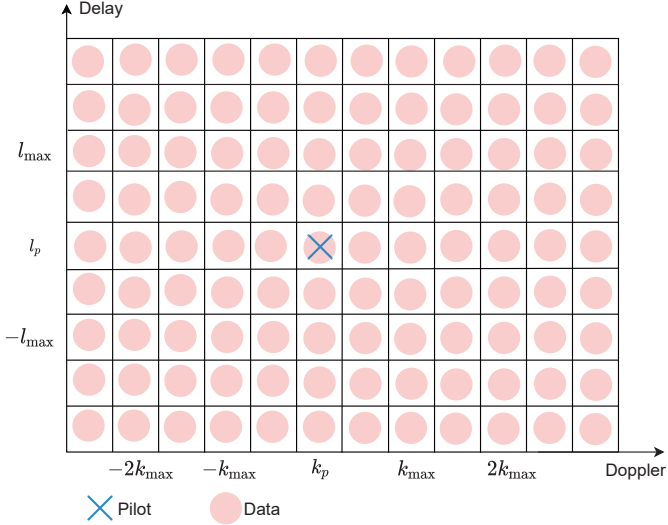}
\caption{{The proposed symbol placement scheme in a single OTFS frame.}}%
\label{new_scheme}%
\end{figure}

As for the considered ISAC-assisted OTFS modulation, given the known delays and Dopplers associated with all paths, we easily determine the corresponding delay and Doppler indices, i.e., $\bar{l}_{i,p}$ and $\bar{k}_{i,p}$. The prediction of the interference pattern based on the sensing parameters enables us to adopt a guard space free symbol placement scheme, given by
\begin{align}
	x_i^{\rm U}\left[k,l\right] =\left\{\begin{array}{ll}
x_{\rm pl},& k = k_{\rm pl}, l = l_{\rm pl},\\
\textrm{Data symbol}, &\forall k,~l,
\end{array}
\right.
\end{align}
as illustrated in Fig. \ref{new_scheme}. With the new symbol placement scheme, more data symbols can be carried in an OTFS frame at the cost of introducing interference between the pilot and data symbols. Based on pre-obtained $\bar{l}_{i,p}$ and $\bar{k}_{i,p}$, the channel path gain can be estimated as
\begin{align}
\hat{h}_{i,p} = \frac{y_i^{\rm U}\left[k_p+\bar{k}_{i,p},l_p+\bar{l}_{i,p}\right]}{x_{\rm pl}}.	
\end{align}
Obviously, as the received sample $y_i^{\rm U}\left[k_p+\bar{k}_{i,p},l_p+\bar{l}_{i,p}\right]$ is also contributed by the data symbols, the estimate $\hat{h}_{i,p}$ is with uncertainty. To quantify the uncertainty, we reconsider the uplink input-output relationship \eqref{io_dd_up} that a received sample is contributed by a total number of $P$ symbols. Therefore, for any received sample $y_i^{\rm U}\left[k_p+\bar{k}_{i,p},l_p+\bar{l}_{i,p}\right]$that is contributed by the pilot, we have
\begin{align}\label{io_ce}
	y_i^{\rm U}\left[k_p+\bar{k}_{i,p},l_p+\bar{l}_{i,p}\right] = h_{i,p} x_{\rm pl} + I\left[k,l\right],
\end{align}
where $I\left[k,l\right]$ denotes the interference term induced by both the data symbols and the noise. Hence, the uncertainty is determined by
\begin{align}
\sigma^2_{i,p} &= \mathbb{E}\left\{\left|\frac{I\left[k,l\right]}{x_{\rm pl}}\right|^2\right\}\nonumber\\
& = \frac{P\cdot E_s+N_0}{E_p}\approx P \frac{E_s}{E_p},
\end{align}
where $E_p$ denotes the power of pilot symbol. Obviously, a higher ratio of pilot-to-data power will lead to a smaller uncertainty of the channel estimation result.

\subsubsection{Data Detection for Uplink Transmission}
Having obtained the estimates of the uplink channel, the RSU can extract the carried information from the uplink transmission block. Let us denote $\tb{y}_i^{\rm U}$ as the vector containing all uplink received samples corresponding to vehicle $i$, the symbol-wise maximum \emph{a posteriori} (MAP) detector is given by
\begin{align}
	\hat{x}_i^{\rm U} \left[k,l\right] = \arg\max_{x_i^{\rm U}\in\mathcal{A}} p\left({x}_i^{\rm U} \left[k,l\right]|\tb{y}_i^{\rm U}\right),
\end{align}
where $p\left({x}_i^{\rm U} \left[k,l\right]|\tb{y}_i^{\rm U}\right)$ is the marginal distribution obtained by marginalizing the joint distribution
\begin{align}
	p\left({x}_i^{\rm U} \left[k,l\right]|\tb{y}_i^{\rm U}\right) = \sum_{\sim {x}_i^{\rm U} \left[k,l\right]} p\left(\tb{x}_i^{\rm U}|\tb{y}_i^{\rm U}\right).
\end{align}
For efficiently calculating the marginal distribution, we resort to the Bayes theorem such that the joint distribution can be factorized as
\begin{align}
	p\left(\tb{x}_i^{\rm U}|\tb{y}_i^{\rm U}\right) = p\left(\tb{x}_i^{\rm U}\right)p\left(\tb{y}_i^{\rm U}|\tb{x}_i^{\rm U}\right).
\end{align}

\begin{figure}[!t]
\centering
\includegraphics[width=.5\textwidth]{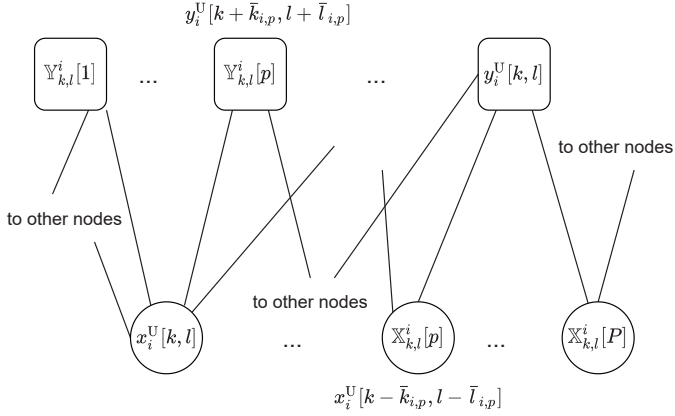}
\caption{{Factor graph representation for the uplink symbol detection.}}%
\label{FG_OTFS}%
\end{figure}

As each symbol experiences a $P$-path channel, there are $P$ received samples related to ${x}_i^{\rm U} \left[k,l\right]$. We denote the set containing all received samples related to ${x}_i^{\rm U} \left[k,l\right]$ as $\mathbb{Y}_{k,l}^{i}$, whose $p$-th element $\mathbb{Y}_{k,l}^{i}[p]=y_i^{\rm U}\left[k+\bar{k}_{i,p},l+\bar{l}_{i,p}\right]$. Similarly,  any received sample $y_i^{\rm U}\left[k,l\right]$ is related to $P$ symbols. Hence, we can define $\mathcal{X}_{k,l}^{i}$ as the set of all symbols contributed to $y_i^{\rm U}\left[k,l\right]$. The $p$-th element in $\mathcal{X}_{k,l}^{i}$ is given by $\mathcal{X}_{k,l}^{i}[p]=x_i^{\rm U}\left[k-\bar{k}_{i,p},l-\bar{l}_{i,p}\right]$. Given the independent assumption of the noise terms associated with the received samples, the likelihood function $p\left(\tb{y}_i^{\rm U}|\tb{x}_i^{\rm U}\right)$ can be factorized as
\begin{align}\label{factorization}
	p\left(\tb{y}_i^{\rm U}|\tb{x}_i^{\rm U}\right) = \prod_{k,l} p\left(y_{i}^{\rm U}\left[k,l\right]|\mathcal{X}_{k,l}^{i}\right).
\end{align}
In particular, $p\left(y_{i}^{\rm U}\left[k,l\right]|\mathcal{X}_{k,l}^{i}\right)$ has a Gaussian representation as
\begin{align}\label{likeli}
	&p\left(y_{i}^{\rm U}\left[k,l\right]|\mathcal{X}_{k,l}^{i}\right)\\
	&\propto \exp\left(-\frac{\left|y_{i}^{\rm U}\left[k,l\right]-\sum_{{p=1}}^P \tb{g}_{p}^{\rm H} \tb{b}(\theta_{i,p}) h_{i,p} \mathcal{X}_{k,l}^{i}[p]\right|^2}{N_0}\right),\nonumber
\end{align}
where we assume that the term $e^{-j2\pi \nu_{i,p}\tau_{i,p}}$ has been compensated using the predicted delay and Doppler. For notational brevity, we further define $\kappa_{i,p}=\tb{g}_p^{\rm H}\tb{b}(\theta_{i,p})$ as the receive beamforming gain. For the \emph{a priori} distribution $p\left(\tb{x}_i^{\rm U}\right)$, as the transmitted symbols are independent, $p\left(\tb{x}_i^{\rm U}\right)$ can be fully factorized as
\begin{align}\label{apriori}
p\left(\tb{x}_i^{\rm U}\right) = \prod_{k,l} p\left({x}_i^{\rm U}\left[k,l\right]\right).
\end{align}
For coded systems, the \emph{a priori} distribution $p\left({x}_i^{\rm U}\left[k,l\right]\right)$ is calculated based on the output log-likelihood ratio (LLR) from the channel decoder. As for uncoded systems, $p\left({x}_i^{\rm U}\left[k,l\right]\right)$ is a discrete distribution with equal probabilities on all constellation points. Based on \eqref{factorization}-\eqref{apriori}, we can represent the the joint distribution by a factor graph \cite{yuan2019iterative}, as depicted in Fig. \ref{FG_OTFS}.

Relying on the factor graph, we can calculate the messages passing on the edges following the sum-product algorithm (SPA) \cite{kschischang2001factor}. For notational purpose, we use $\mu$ to denote a message from a variable node to a function node and $\psi$ to denote a message from a function node to a variable node. In particular, the message from a variable node $x_i^{\rm U}\left[k,l\right]$ to its connected function node $\mathbb{Y}_{k,l}^i[p]$ is given by
\begin{align}
\mu(x_i^{\rm U}\left[k,l\right] \to \mathbb{Y}_{k,l}^i[p]) = \prod_{\substack{p'=1\\p'\neq p}}^P \psi(\mathbb{Y}_{k,l}^i[p']\to x_i^{\rm U}\left[k,l\right]),	
\end{align}
while the message from a function node $y_i^{\rm U}\left[k,l\right]$ to a variable node $\mathcal{X}_{k,l}^i [p]$ is expressed as
\begin{align}\label{psiytox}
	\psi&(y_i^{\rm U}\left[k,l\right]\to \mathcal{X}_{k,l}^i [p])  \\&=\sum_{\mathcal{X}_{k,l}^i\sim p} p\left(y_i^{\rm U}\left[k,l\right]|\mathcal{X}_{k,l}^i\right)	\prod_{\substack{p'=1\\p'\neq p}}^P\mu(\mathcal{X}_{k,l}^i[p']\to y_i^{\rm U}\left[k,l\right]).\nonumber
\end{align}
Given $\chi_q\in\mathcal{A}$ and $p_q$ being the $q$-th constellation point and associated probability, $\mu(\mathcal{X}_{k,l}^i[p]\to y_i^{\rm U}\left[k,l\right])$ is expected to have the form of
\begin{align}
	\mu(\mathcal{X}_{k,l}^i[p]\to y_i^{\rm U}\left[k,l\right]) =\sum_{q=1}^{|\mathcal{A}|} p_{q}\delta\left(\mathcal{X}_{k,l}^i[p]-\chi_q\right).
\end{align}
By substituting \eqref{likeli}, eq. \eqref{psiytox} can be obtained in a Gaussian form with mean
\begin{align}
	\mathbb{E}&\left[y_i^{\rm U}\left[k,l\right]\to \mathcal{X}_{k,l}^i [p]\right] \\ &=\frac{y_{i}^{\rm U}\left[k,l\right]-\sum_{p'=1,\neq p}^P \kappa_{i,p'}\hat{h}_{i,p'} \mathbb{E}\left[\mathcal{X}_{k,l}^i[p]\to y_i^{\rm U}\left[k,l\right]\right]}{\kappa_{i,p}\hat{h}_{i,p}},\nonumber
\end{align}
and variance
\begin{align}\label{vytox}
		\mathbb{V}&\left[y_i^{\rm U}\left[k,l\right]\to \mathcal{X}_{k,l}^i [p]\right] = \frac{1}{|\kappa_{i,p}\hat{h}_{i,p}|^2}\cdot \Big(N_0\\&+\sum_{\substack{p'=1,p'\neq p}}^P |\kappa_{i,p'}\hat{h}_{i,p'}|^2\cdot \mathbb{V}\left[\mathcal{X}_{k,l}^i[p']\to y_i^{\rm U}\left[k,l\right]\right]\Big),\nonumber
\end{align}
which denote the information of $\mathcal{X}_{k,l}^i [p]$ from $y_i^{\rm U}\left[k,l\right]$,
where
\begin{align}
	\mathbb{E}\left[\mathcal{X}_{k,l}^i[p]\to y_i^{\rm U}\left[k,l\right]\right] &= \sum_{q} p_q \chi_q~\textrm{and}\\
	\mathbb{V}\left[\mathcal{X}_{k,l}^i[p]\to y_i^{\rm U}\left[k,l\right]\right] &= \sum_q p_q |\chi_q|^2-\left|\sum_{q} p_q \chi_q\right|^2
\end{align}
denote the mean and variance of the information of $\mathcal{X}_{k,l}^i[p]$ that contributes to the received sample $y_i^{\rm U}\left[k,l\right]$, respectively.
Note that when calculating \eqref{vytox}, we adopt the estimated channel $\hat{h}_{i,p'}$. As discussed in Sec. \ref{subsubsec1}, the channel estimation is with an uncertainty of $\sigma_{i,p}^2$. Ignoring the uncertainty would lead to performance loss. Therefore we take into account the uncertainty $\sigma_{i,p}^2$ for message derivation. According to the Mellin transformation \cite{oberhettinger2012tables}, the variance of the product of two independent variables $x_1$ and $x_2$ is given by
\begin{align}
	\mathbb{V}&\left[x_1x_2\right] \nonumber\\&= \left(|\mathbb{E}[x_1]|^2+\mathbb{V}[x_1]\right)\left(|\mathbb{E}[x_2]|^2+\mathbb{V}[x_2]\right)
-|\mathbb{E}[x_1]\mathbb{E}[x_2]|^2 \nonumber\\&= \mathbb{V}[x_1]\mathbb{V}[x_2]+|\mathbb{E}[x_2]|^2\mathbb{V}[x_1]+|\mathbb{E}[x_1]|^2\mathbb{V}[x_2].
\end{align}
As the channel tap $h_{i,p}$ and transmitted symbol $\mathcal{X}_{k,l}^i\left[p\right]$ are independent, by taking into account the channel estimation uncertainty, we can revise \eqref{vytox} as
\begin{align}\label{vytox1}
		&\mathbb{V}\left[y_i^{\rm U}\left[k,l\right]\to \mathcal{X}_{k,l}^i [p]\right] = \frac{1}{|\kappa_{i,p}\hat{h}_{i,p}|^2}\cdot \bigg(N_0+\sum_{\substack{p'=1,\\p'\neq p}}^P E_s \Big(\sigma_{i,p}^2 \nonumber
\\& +\left(|\kappa_{i,p'}\hat{h}_{i,p'}|^2+\sigma_{i,p}^2\right)\cdot \mathbb{V}\left[\mathcal{X}_{k,l}^i[p']\to y_i^{\rm U}\left[k,l\right]\right]\Big)\bigg).\end{align}
In \eqref{vytox1}, the information of channel uncertainty is also included, which helps to improve the detection performance.

To calculate the probability associated with each constellation point for $\psi(\mathbb{Y}_{k,l}^i [p]\to x_i^{\rm U}\left[k,l\right])$, we substitute $x_i^{\rm U}\left[k,l\right]=\chi_q$ into \eqref{psiytox} and obtain the probability corresponding to the $q$-th constellation point, denoted by $p^{k,l}_{q}[p]$. Then we are able to update the messages from variable nodes to function nodes. Finally, the approximation for the marginal distribution of $x_i^{\rm U}\left[k,l\right]$ is given by
\begin{align}
	p\left({x}_i^{\rm U} \left[k,l\right]|\tb{y}_i^{\rm U}\right) &= \prod_p \psi(\mathbb{Y}_{k,l}^i[p]\to x_i^{\rm U}\left[k,l\right])\nonumber\\
	&=\sum_{q=1}^{|\mathcal{A}|} p^{k,l}_q \delta\left(x_i^{\rm U}\left[k,l\right]-\chi_q\right),
\end{align}
where the probability $p^{k,l}_q$ is given by
\begin{align}
p^{k,l}_q = \frac{\prod_p p^{k,l}_{q}[p]}{\sum_q p^{k,l}_q}.
\end{align}
On the other hand, for coded systems, the probabilities $p^{k,l}_q,~\forall q$, are used to calculate the LLRs of the coded bits, which are fed to the channel decoder for extracting the original information bits. For the uncoded system, the data symbol is detected by simply comparing the probabilities $p^{k,l}_q$ and set $\hat{x}_i^{\rm U} \left[k,l\right] = \chi_q$ for the largest $p^{k,l}_q$, $\forall q$.


\section{Simulation Results}
We consider the vehicular network illustrated in Fig. 1. We assume that there are four vehicles moving on a two-lane road. Without loss of generality, we assume that the RSU is located at location $[0,0]^{\rm T}$. The coordinates of the vehicles are initialized at $[-30,50]^{\rm T}$, $[20,50]^{\rm T}$, $[5,20]^{\rm T}$, and $[40,20]^{\rm T}$. The vehicle speeds are uniformly and randomly drawn from $10\leq |s_i| \leq 15$ m/s, $\forall i$. The operating carrier frequency for the RSU is $f_c=3$ GHz and the subcarrier spacing is $6$ kHz. Consequently, the maximum Doppler index is $k_{\rm max}=3$ for downlink transmission and $k_{\rm max}=6$ for uplink transmission\footnote{For uplink transmission, two vehicles may move towards the opposite direction and thus the maximum Doppler index is doubled.}. The maximum delay index is determined by the maximum range between the vehicles and the RSU, which is roughly set it as $l_{\rm max}=10$. Moreover, We consider $200$ time instants and the duration for each time instant is $\Delta T=0.02~$s. For the noise term corresponding to the echoes, we assume that $\sigma^2=1$ \cite{9171304}. The size of the OTFS frame is set to $M=128$ and $N=30$. Finally, the binary phase shift keying (BPSK) modulation is used for symbol mapping.

 \begin{figure}[!t]
\centering
\includegraphics[width=.5\textwidth]{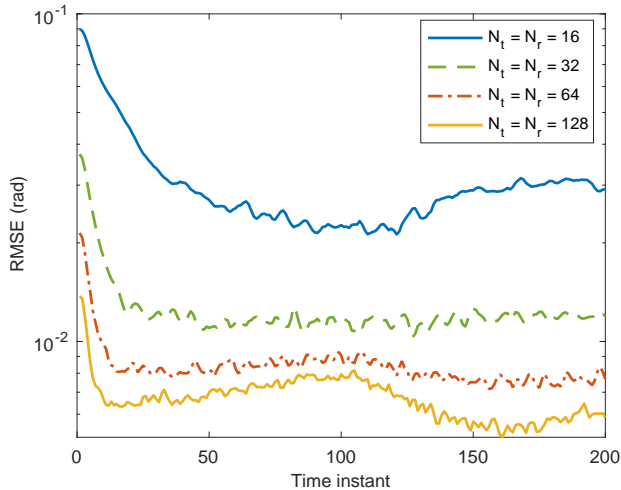}
\caption{{The RMSE of the angle estimate versus the time index.}}%
\label{fig1}%
\end{figure}

We first consider the sensing performance of the vehicle states based on the OTFS-ISAC signal. In Fig. \ref{fig1}, we depict the root mean squared error (RMSE) of the angle estimate for all four vehicles versus the time index. Four configurations with different sizes of the antenna array, i.e., $N_t=N_r=16$, $N_t=N_r=32$, $N_t=N_r=64$, $N_t=N_r=128$ are considered. We can observe that for all configurations, the RMSE of the angle estimate sharply decreases with the time index in the first few time instants. This is because more information can be gleaned from the state evolution of the vehicles, which improves the estimation accuracy. Moreover, the larger number of antennas, the better estimation performance is achieved, since the higher antenna array gain leads to a higher SNR. Obviously, adopting a larger antenna array indicates a narrower beamwidth, which imposes challenges on accurate beam alignment. Nevertheless, the RMSE can still achieve the level of $10^{-2}$ rad, showing the effectiveness of the proposed beam alignment algorithm.

 \begin{figure}[!t]
\centering
\includegraphics[width=.5\textwidth]{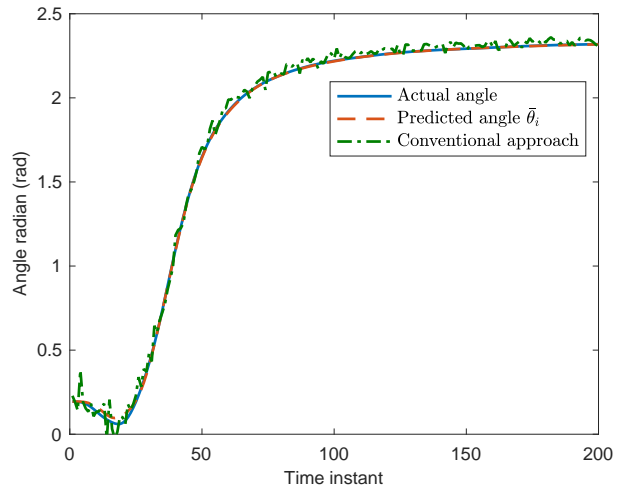}
\caption{{The angle tracking performance versus the time instant.}}%
\label{fig2}%
\end{figure}

In Fig. \ref{fig2}, we show the tracking performance of the angle of the third vehicle with initial coordinate $[5,20]^{\rm T}$ relative to the RSU based on the proposed algorithm \eqref{ML_radar} and the feedback-based beam pairing scheme \cite{shaham2019fast}. The antenna configuration is $N_t=N_r=64$. In particular, the conventional beam pairing approach adopts one pilot symbol for pairing the beam directions before the data transmission. As discussed in Sec. III-A, the feedback-based scheme incurs considerable signaling latency as well as estimation overhead. Moreover, by adopting OTFS-ISAC signal for downlink transmission, the whole downlink block can be used as pilots for angle estimation, which will provide a high matched filter gain. Observed from Fig. \ref{fig2}, we can see that benefiting from the matched filter gain $G_m$, the predicted angle $\bar{\theta}_i$ based on \eqref{predicted_angle} can accurately track the actual angle with a negligible error, which validates our motivation of predicting the angle for formulating the transmit beamformers at the RSU. Compared with the proposed approach, tracking error can be observe for the conventional feedback-based scheme using only one pilot. Considering that the beamwidth is narrow for massive MIMO scenario, the tracking error of angle will result in the misalignment of the transmit beams for the conventional scheme, which will lead to a degraded receive SNR.

 \begin{figure}[!t]
\centering
\includegraphics[width=.5\textwidth]{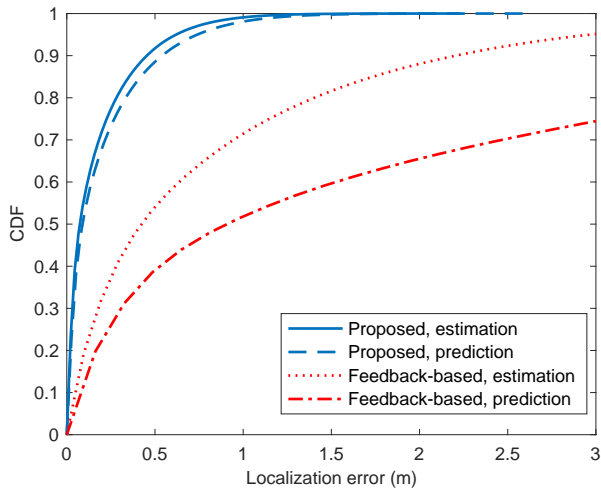}
\caption{{The CDF of the localization error for different algorithms.}}%
\label{fig3}%
\end{figure}

In Fig. \ref{fig3}, the cumulative distribution function (CDF) of the localization error is provided. The curves for both the location estimate $\hat{\tb{q}}_i$ and the predicted location $\bar{\tb{q}}_i$ are illustrated. For comparison, the estimation and prediction results based on the feedback-based approach are included in Fig. \ref{fig3}. We can see that by exploiting the temporal relationship of the vehicle states in two consecutive time instants, the proposed algorithm can efficiently locate the vehicles. In contrast, due to the low SNR gain and relatively high angle estimation error, there is a remarkable localization error for the feedback-based approach.

\begin{figure}[!t]
\centering
\includegraphics[width=.5\textwidth]{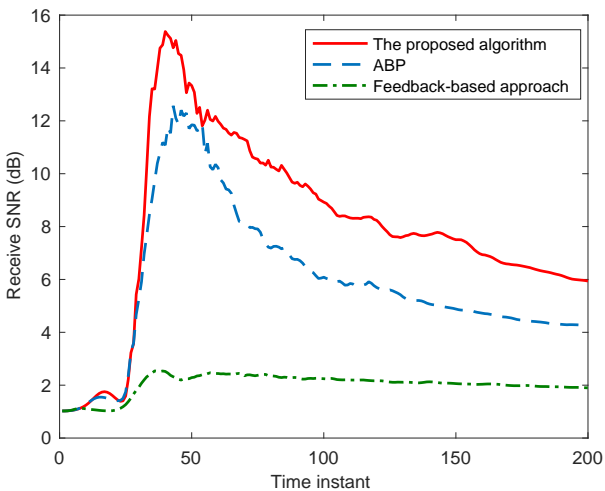}
\caption{{The receive SNR at vehicle $3$ versus time index.}}%
\label{fig4}%
\end{figure}

The communication performance relies on the receive SNR. In Fig. \ref{fig4}, we compare the receive SNR at the third vehicle based on the proposed algorithm and two benchmark algorithms, i.e., the feedback-based method and the auxiliary beam pairing (ABP) approach in \cite{7472306}, to verify the effectiveness of the proposed prediction-based beam alignment scheme. The number of transmit antenna is $N_t = 64$ and the transmit SNR is normalized to $0$ dB to emphasize the SNR gain. We can see that the proposed algorithm significantly outperforms both of the benchmark methods. The receive SNR increases in the first few time instants and then drops rapidly, which is due to that the vehicle first moves towards the RSU and then moves away from it. The feedback-based method is not capable of tracking the variation of the angle using only one pilot and therefore results in an SNR loss. As for the ABP approach, although it is easy to be implemented, the fixed search space for beam directions makes it suffer from a slight SNR degradation.

 \begin{figure}[!t]
\centering
\includegraphics[width=.5\textwidth]{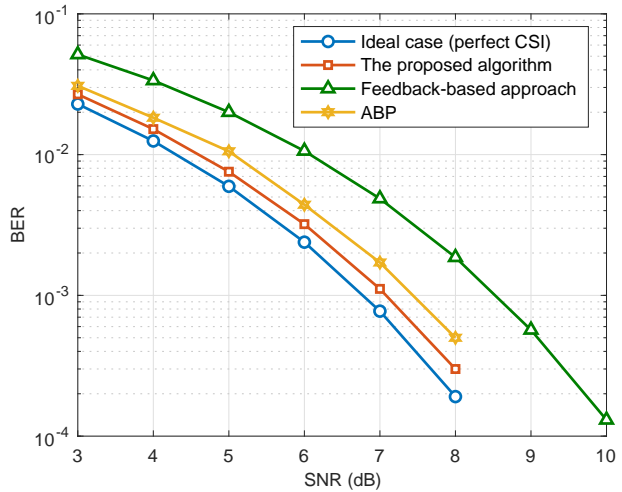}
\caption{{The BER performance of different algorithms for uplink transmission.}}%
\label{fig5}%
\end{figure}
Next, we move our focus to the communication performance. We first consider the downlink transmission. We show the bit-error-rate (BER) performance versus the receive SNR of the proposed algorithm in Fig. \ref{fig5}. The BER performance of two benchmark schemes using the feedback-based and ABP beam alignment schemes followed by the classic OTFS channel estimation \cite{8671740} and the performance corresponding to perfect channel state information (CSI) are depicted in Fig. \ref{fig5} as well. We can observe that the performance for the proposed algorithm can approach that of the ideal case with a perfectly known channel. For the benchmark scheme using conventional beam pairing and channel estimation algorithm, although the channel can be accurately estimated, the degraded receive SNR would lead to a BER performance loss. Note that the downlink communication efficiency can be improved by the proposed algorithm since no pilots are required. Thus, Fig. \ref{fig5} shows the superiority of using OTFS-ISAC signal for downlink transmission.

\begin{figure}[!t]
\centering
\includegraphics[width=.5\textwidth]{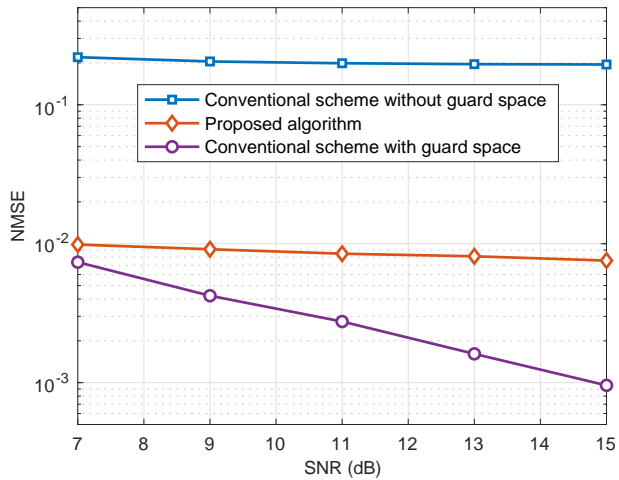}
\caption{{The NMSE of channel estimation for the proposed algorithm and the conventional scheme.}}%
\label{fig6}%
\end{figure}

For uplink transmission, Fig. \ref{fig6} illustrates the normalized mean square error (NMSE) of the channel estimate based on the superimposed pilot-data symbol placement scheme. The NMSE of channel estimation is given by
\begin{align}
{\rm NMSE}_{h_i} = \frac{\sum_p |\hat{h}_{i,p}-h_{i,p}|^2}{\sum_p |h_{i,p}|^2}.	
\end{align}
The power of the pilot is assumed to be $20$ dB higher than that of the data symbols. The NMSE performances for channel estimation scheme \cite{8671740} with and without guard space are plotted as reference. We can observe that the NMSE of the channel estimate relying on predicted parameters and the proposed symbol placement scheme is bounded by the uncertainty $\sigma^2_{i,p}$, which is approximately $10^{-2}$ for the aforementioned pilot power. The conventional channel estimation approach with guard space can provide the best performance as there exists no interference between the pilot and data symbols. If we also adopt the proposed symbol placement scheme, the conventional approach suffers from a significant performance degradation.

\begin{table}[]
\centering
{
\caption{Training Overhead for Channel Estimation}
\label{table1}
\begin{tabular}{|l|c|c|}
\hline
Placement scheme & Number of  & Training Overhead\\
&guard symbols&\\
\hline
Conventional  & $480$ & $12.5$\%\\
\hline
Proposed & $0$ & $0$\%\\
\hline
\end{tabular}
}
\end{table}

Then, we compare the training overhead for channel estimation in Table. II based on the symbol placement scheme shown in Fig. 3 and the proposed one in Fig. 4. A total number of $8k_{\rm max}l_{\rm max} = 480$ grids out of $NM=3840$ DD domain grids are reserved for channel estimation, yielding a training overhead of $12.5$ \%. The overhead will become even higher if the vehicle speed goes higher or the OTFS frame size becomes smaller. In contrast, as shown in Fig. 4, all DD domain grids are used to carry data symbols, indicating the training overhead for the proposed scheme is as low as zero. In addition, the NMSE of channel estimate at the level of $10^{-2}$ has only negligible impact on the BER performance, which will be shown in the following.

\begin{figure}[!t]
\centering
\includegraphics[width=.5\textwidth]{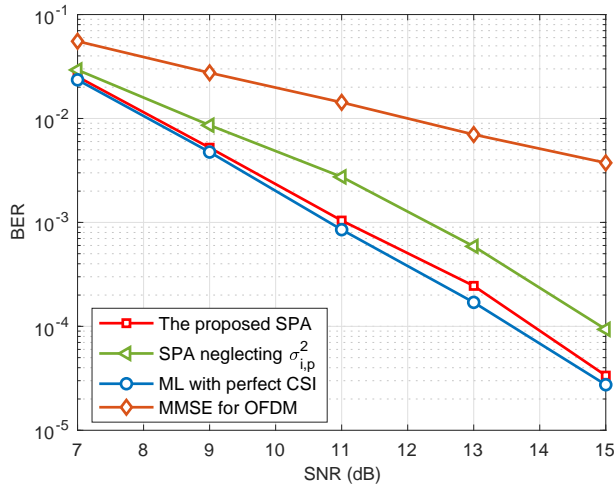}
\caption{{The BER performance of different algorithms for uplink transmission.}}%
\label{fig7}%
\end{figure}

Finally, the BER performance versus SNR for the uplink transmission is depicted in Fig. \ref{fig7}. Four algorithms, i.e., the proposed channel estimation and detection, SPA detection that neglects the uncertainty for channel estimation, symbol-wise ML detection with perfect CSI, and the minimum mean squared error (MMSE) detection for an OFDM system are included for comparison. Since the orthogonality of the subcarriers does not hold in vehicular scenarios, MMSE detection for the OFDM system shows a severe performance degradation. The BER performance of the proposed channel estimation and detection scheme can approach that of the ideal case with perfect CSI and symbol-wise ML detection with the conventional OTFS channel estimation. Moreover, compared with the reference algorithm that neglects the estimation uncertainty, the proposed algorithm has shown a BER performance improvement. Through the comparison, we show the OTFS-ISAC system relying on the proposed channel estimation and detection method can reliably support the uplink transmission in vehicular networks.

\section{Conclusions}
In this paper, we proposed a novel ISAC-assisted OTFS transmission scheme for vehicular networks. Based on the echoes reflected by the vehicles, the motion parameters of vehicles can be estimated at the RSU, which are exploited for predicting the vehicle states, e.g., location and speed in the following time instant. Consequently, benefiting from the slow time-varying DD domain channel coefficients, the channel parameters can be predicted. For the downlink transmission, neither dedicated pilots for estimating the downlink channel parameters nor for beam paring are required. As for the uplink transmission, relying on the predicted delays and Dopplers, we developed a guard space free symbol placement scheme, which yields a much lower channel estimation training overhead compared to the conventional approach. Simulation results showed that the proposed ISAC-assisted OTFS transmission scheme can reduce the required training overhead while providing reliable communications.
\bibliographystyle{IEEEtran}
\bibliography{v2x}
\end{document}